\def\nfn{\nu F_{\nu}}
\def\nfnsy{\left( \nfn \right)_{\rm sy}}
\def\nfnssc{\left( \nfn \right)_{\rm SSC}}
\def\nfnerc{\left( \nfn \right)_{\rm ERC}}
\def\nfnsyn{F_{{\rm sy}, -10}}
\def\nfnsscn{F_{{\rm SSC}, -10}}
\def\nfnercn{F_{{\rm ERC}, -10}}
\def\esy{\epsilon_{\rm sy}}
\def\essc{\epsilon_{\rm SSC}}
\def\eerc{\epsilon_{\rm ERC}}
\def\estar{\epsilon_{\ast}}
\def\esyn{\epsilon_{{\rm sy}, -7}}
\def\esscn{\epsilon_{{\rm SSC}, -1}}
\def\eercn{\epsilon_{{\rm ERC}, 2}}
\def\estarn{\epsilon_{\ast, -5}}
\def\eb{\epsilon_B}
\def\gcr{\gamma_{\rm cr}}
\def\fsp{f_{\rm sp}}
\def\fsy{f_{\rm sy}}
\def\ferc{f_{\rm ERC}}
\def\fssc{f_{\rm SSC}}
\def\uext{u_{\rm ext}}

\def\taur{\tau_{\rm repr}}
\def\ls{\lower4pt\hbox{${\buildrel < \over \sim}$}}
\def\gs{\lower4pt\hbox{${\buildrel > \over \sim}$}}

\documentclass[12pt, preprint]{aastex}

\slugcomment{Submitted to {\it The Astrophysical Journal Letters} 
on 06/21/01 \\
Revised: 07/30/01}

\shorttitle{An evolutionary scenario for AGN unification}
\shortauthors{B\"ottcher \& Dermer}

\begin{document}

\title{An Evolutionary Scenario for Blazar Unification}

\author{M. B\"ottcher\footnote{Chandra Fellow}}
\affil{Department of Physics and Astronomy, Rice University, MS 108, \\
6100 S. Man Street, Houston, TX 77005 - 1892, USA}
\email{mboett@spacsun.rice.edu}

\and

\author{C. D. Dermer}
\affil{E. O. Hulburt Center for Space Research, Code 7653,\\
Naval Research Laboratory, Washington, DC 20375-5352}
\email{dermer@gamma.nrl.navy.mil}

\begin{abstract}
Blazar subclasses ranging from flat-spectrum radio quasars (FSRQs)
through low-frequency-peaked BL Lac objects (LBLs) to
high-frequency-peaked BL Lac objects (HBLs) exhibit a sequence of
increasing spectral hardness with decreasing luminosity that cannot be
explained solely by orientation effects. Using an analytic model for
the synchrotron, synchrotron-self-Compton, and Compton-scattered
external radiation from blazar jets, we propose an evolutionary
scenario that links these blazar subclasses in terms of a reduction of
the black-hole accretion power with time. As the circumnuclear
material accretes to fuel the central engine, less gas and dust is
left to scatter accretion-disk radiation and produce an external
Compton-scattered component in blazar spectra.  This evolutionary
trend produces the sequence FSRQ $\rightarrow$ LBL $\rightarrow$ HBL.
Such a scenario may also link radio-loud AGNs with ultraluminous
infrared galaxies and optical QSOs, if the latter constitute the high
Eddington-ratio epoch of supermassive black hole growth, as suggested
by the observed anti-correlation between radio and soft X-ray activity
in some Galactic black-hole candidates.
\end{abstract}


\section{Introduction}

During the past decade, significant progress has been made in
understanding the physics of active galactic nuclei (AGN) and the
connection between different subclasses of AGN. In particular, various
lines of evidence indicate that FR II radio galaxies are the parent
population of radio-loud quasars, with the relativistic jets of FSRQs
being closely aligned to our line-of-sight (for a review, see
\cite{up95}). Orientation effects are also thought to unify FR I radio
galaxies with BL Lac objects. The BL-Lac/FR-I subclasses have lower
average luminosities than the FSRQ/FR-II subclasses. Among
blazars, which include highly polarized and optically violently
variable quasars, FSRQs and BL Lac objects, there appears to be an
almost continuous sequence of properties from FSRQs through LBLs to
HBLs. This trend is characterized by decreasing bolometric
luminosities, a shift of the peak frequencies of their broadband
spectral components towards higher values, and a decreasing fraction
of power in $\gamma$ rays compared with lower-frequency radiation
\citep{sambruna96,fossati98}.

In the framework of relativistic jet models, the low-frequency (radio
-- optical/UV) emission from blazars is interpreted as synchrotron
emission from nonthermal electrons in a relativistic jet. The
high-frequency (X-ray -- $\gamma$-ray) emission is thought to be
produced via Compton upscattering of low frequency radiation by the
same electrons responsible for the synchrotron emission. Possible
sources of soft seed photons for Compton scattering are the
synchrotron photons themselves
\citep{maraschi92,bm96} or external photons, presumably dominated
by the accretion disk emission which can enter the jet either
directly \citep{dsm92,ds93} or after reprocessing by circumnuclear
gas and dust \citep{sikora94,dss97}. In addition, a significant 
contribution to the soft radiation field may also be provided 
by infrared emission from dust in the vicinity of the AGN
\citep{blaz00,arbeiter01}. 

The blazar sequence has been studied by \cite{ghisellini98} using a 
large sample of blazar broadband spectra. They suggest that 
along the sequence HBL $\to$ LBL $\to$ FSRQ, an increasing energy
density of the external radiation field leads to an increasing amount
of Compton cooling. As a result, the maximum energy in the electron
distribution decreases, causing the synchrotron and Compton peaks to
shift to lower frequencies. This explanation for the blazar sequence
was developed further by \cite{geo01}, who argue that the radiating
jet plasma is outside the broad-line scattering region in weak sources
and within it in powerful sources.  Detailed model fits to the
broadband spectra of several blazars also indicate that a decreasing
contribution of the external radiation to the seed photon field
reproduces the FSRQ $\to$ LBL $\to$ HBL sequence.  While FSRQs
generally require a dominant external Compton contribution to produce
the observed $\gamma$-ray emission (e.g.,
\cite{sambruna97,muk99,hartman01}), HBLs can be successfully fitted
with pure synchrotron self-Compton (SSC) models (e.g.,
\cite{mk97,pian98,petry00}). The LBL BL~Lacertae appears 
to be intermediate between those two classes, requiring a
non-negligible contribution from external Comptonization to reproduce
its $\gamma$-ray spectrum \citep{madejski99,bb00}. Orientation effects
certainly play a role in spectral variations among sources within a
given subclass, or even to account for the blue blazar subclass
\citep{geo00}, but cannot explain the full range of blazar properties
\citep{sambruna96}.

It is uncertain whether the different blazar types are connected
through an evolutionary sequence, or whether they constitute limited
periods of AGN activity on parallel evolutionary paths. 
An evolutionary sequence has recently been proposed by \cite{delia00} and 
\cite{cavaliere01} who consider the efficiency of energy extraction 
from the accretion flow onto the central supermassive black hole
through magnetic fields anchored in an optically thick accretion 
disk vs. energy extraction from rotational energy of the black 
hole via the Blandford-Znajek (BZ) mechanism \citep{bz77}. They 
argue that BL Lac objects can plausibly be powered by the BZ
mechanism without a significant contribution from an accretion
disk, while FSRQs require a dominant contribution from the
accretion disk to power the relativistic jets responsible for
the production of the nonthermal broadband spectra. They suggest
that a gradual depletion of the central environment in FSRQs by
accretion onto the central black hole may be responsible for 
a transition from the disk-powered jet production (in FSRQs) 
to the BZ-powered jet production mode (in BL Lac objects).
Furthermore, they find that a cosmological BL Lac birth rate 
proportional to the rate of ``deaths'' of FSRQs (through
depletion of their central environments) is consistent with
the cosmological evolution observed for both quasars and
BL Lac objects. The masses of the central black holes in 
radio quasars generally exceed $\sim 3\times 10^8 \, M_{\odot}$ 
\citep{wandel99,laor00}. Similarly detailed mass determinations 
have so far not been possible for the lineless BL Lacertae objects. 
The cluster environments of FR II radio galaxies and radio quasars 
on the one hand, and FR I radio galaxies and BL Lacs on the other 
hand appear consistent with an evolution of the former into the 
latter class of objects \citep{hl91,ye93,up95}. Also, the 
distribution of central black-hole masses in nearby non-active 
galaxies appears to extend to larger values than the distribution 
of radio-quasar black-hole masses \citep{laor00}, consistent with 
some fraction of present-day giant elliptical galaxies hosting the 
remnants of earlier quasar activity.

Here we expand on the suggestion of a cosmological evolutionary
sequence \citep{delia00,cavaliere01} from FSRQs to BL~Lac objects by
demonstrating that such a sequence plausibly and self-consistently
explains the trend in their spectral properties. We employ a combined 
SSC/external Compton model for relativistic jet emission, as described 
in \S \ref{modelsetup}, and derive parameter values from the observed 
multifrequency signatures of FSRQs and HBLs. The evolutionary sequence 
connecting the respective parameter regimes is presented in \S 
\ref{sequence}. We discuss our results and suggest a possible 
evolutionary connection between radio-loud and radio-quiet AGN 
classes in \S \ref{discussion}.

\section{\label{modelsetup}Blazar Model and Parameter Estimation}

The radiating plasma is modeled by a spherical emission region of
co-moving radius $R_b$ that moves outward along the axis of the jet
with bulk Lorentz factor $\Gamma = (1 - \beta_{\Gamma}^2)^{-1/2}$. 
Electrons are injected into the plasma blob with a comoving power-law 
distribution $Q(\gamma) = Q_0 \, \gamma^{-s}$, for electron Lorentz 
factors in the range $\gamma_1 \le \gamma \le \gamma_2$. The jet 
power $L_j$ corresponding to the energy input of injected particles 
into the jet is assumed to be proportional to the accretion disk 
power $L_D$, as suggested by the observed proportionality between 
kinetic jet power and narrow line emission luminosity in radio-loud
AGN reported by \cite{rs91} and the proportionality between core 
radio luminosity and broad emission line luminosity in radio-loud 
AGN found by \cite{celotti97}. In addition, \cite{xu99} have found 
a similar proportionality between the 5~GHz radio luminosity and 
the [O III] 5007 emission line luminosities both for radio-loud 
and radio-quiet AGN. The jet power is related to the injected 
particle distribution through $L_j = f_j \, L_D = m_e c^2 
\int_{\gamma_1}^{\gamma_2} d\gamma \;\gamma \;Q(\gamma)$.

Electrons lose energy through adiabatic expansion, which is treated in
terms of an escape time scale $t_{\rm esc} \equiv \eta \, R_b / c$
with $\eta \approx 1$, and through radiative losses. These losses are
generally dominated by synchrotron and Thomson processes, so they can
be parameterized by the functional form $\dot\gamma_{\rm rad} = -
\nu_0 \, \gamma^2$, where $\nu_0 = 4 c \sigma_T \, \left( u_B + u_{\rm
sy} + u_{\rm ext} \right)/3m_e c^2$.  The terms $u_B$, $u_{\rm sy}$,
and $u_{\rm ext}$ represent the comoving magnetic field, synchrotron,
and external radiation field energy densities, respectively.  If
$u_{\rm ext}$ is dominated by reprocessed accretion-disk radiation due
to circumnuclear gas and dust with a scattering optical depth $\taur$
at a characteristic distance $R_{\rm sc}$ from the central source,
then $u_{\rm ext} \approx \Gamma^2 \, L_D \, \taur / (4 \pi \, R_{\rm
sc}^2 c)$. Our approach is similar to the one used by \cite{ghisellini98},
but differs from their model by our explicitly accounting for adiabatic
losses and escape. These processes lead to the formation of a low-energy 
cutoff in the electron spectrum and force us to consider two distinct
radiative regimes of electron cooling and of the equilibrium electron 
spectra (see below), while the \cite{ghisellini98} approach always
leads to a $\gamma^{-2}$ electron spectrum for $1 \le \gamma \le 
\gamma_1$. We neglect pair processes, which generally turn out to 
be unimportant in the course of detailed spectral modeling of blazars.

Electrons injected with $\gamma \gg \gamma_{\rm cr} \equiv c / (R_b \,
\nu_0)$ cool rapidly prior to escape, whereas electrons injected with
$\gamma \ll \gamma_{cr}$ lose only a small fraction of their energy
before they escape. The competition between electron injection,
cooling and escape establishes an equilibrium distribution given by
\begin{equation}
n_{e,eq} (\gamma) = n_{e, eq}^0 \cases{
(\gamma / \gamma_m)^{-p} & for $\gamma_l \le \gamma \le \gamma_ m$ \cr
(\gamma / \gamma_m)^{-(s + 1)} & for $\gamma_{\rm m} \le \gamma \le 
\gamma_2$ \cr}
\label{ne}
\end{equation}
Two distinct radiative regimes can be defined. If $\gamma_1 <
\gamma_{\rm cr}$, most of the injected electrons cool inefficiently
and the electron population is in the slow cooling regime. In this
case, $p = s$, $\gamma_l = \gamma_1$, $\gamma_m = \gamma_{cr}$, and
$n_{e,eq}^0 = Q_0 \, t_{\rm esc} \, \gamma_m^{-s} / V_b$, where $V_b$ 
is the blob volume.  When $\gamma_1 > \gamma_{\rm cr}$, the electron 
population is in the fast cooling regime, $p = 2$, $\gamma_l = 
\gamma_{cr}$, $\gamma_m = \gamma_1$, and $n_{e, eq}^0 = Q_0 \,
\gamma_m^{-(1 + s)} /[V_b \, \nu_0 \, (s - 1)]$.  The synchrotron, 
and external-Compton radiation emitted by these relativistic 
electron populations are evaluated using the formulae given 
by \cite{dss97}. The SSC radiation is calculated using the 
analytic solution of \cite{tmg98}.

In the absence of an external Compton component, we recover 
the SSC model used to fit data from HBLs such as Mrk~501, 
Mrk~421, or PKS~2155-304 (e.g.,
\cite{mk97,pian98,tmg98,bp99,petry00,sambruna00,fossati00,kataoka00}).  
The parameters in an SSC model
are the Doppler factor $D = \left[ \Gamma
\, (1 - \beta_{\Gamma} \cos\theta_{\rm obs}) \right]^{-1}$, 
the magnetic field $B$, $R_b$, $\gamma_1$, $\gamma_2$, the injection
luminosity $L_j$, and the injection spectral index $s$. In principal,
the escape time scale parameter $\eta$ could also be considered a free
parameter, though we assume (see \cite{kataoka00}) that $\eta \approx
1$, i.e., $t_{\rm esc} \approx R_b/c$.  These studies find typical HBL
parameters of $R_b \sim 10^{15}$ -- $10^{16}$~cm, $\gamma_2 \sim
10^6$, $B \sim 0.1$~G, $D \sim$~20 -- 30, and a rather broad range of
injection luminosities $L_j \sim 10^{37}$~ergs~s$^{-1}$ --
$10^{41}$~erg~s$^{-1}$, depending on the various model details and the
individual flares that are modeled.

The addition of the external Compton component in FSRQs complicates
parameter estimation. However, since detailed modeling of blazar
spectra using self-consistent, combined SSC + ERC spectral models is 
generally a rather time-consuming task, simple formulae to achieve 
order-of-magnitude estimates for the relevant model parameters are 
desirable in order to find a convenient starting point for more
detailed modeling efforts. In the following, we derive such estimates, 
although we must caution that some of the observables, in particular 
those pertaining to the SSC component, are rather uncertain, which 
may introduce significant scatter in the derived analytical parameter
estimates. We assume that the X-ray spectra of FSRQs are
dominated by the SSC process, as has been argued by various authors on
the basis of spectral fitting as well as variability considerations
(e.g., \cite{muk99,hartman01,sikora01}), while the $\gamma$-ray
spectrum is dominated by the external Compton component.  The peak
photon energies and peak $\nu F_{\nu}$ fluxes, $\nu F_{\nu} = 
10^{-10} \, F_{-10}$ ergs~cm$^{-2}$~s$^{-1}$, of the Compton 
and synchrotron components are determined from simultaneous
multiwavelength observations. In the following, photon energies are
given in dimensionless units $\epsilon = h \nu / m_e c^2 = 10^n \,
\epsilon_n$, and we define $\epsilon_B \equiv B / B_{\rm cr} = 
2.3 \times 10^{-14} \, B({\rm G})$. The jets in blazars are
generally believed to be oriented close to the line of sight
where the Doppler beaming factor $D \approx \Gamma$. Thus, for 
simplicity, we assume $D = \Gamma$ in the following. The model 
parameters can then be estimated from the observables through 
the expressions
\begin{equation} 
\esy = {D \eb \gcr^2 \over 1 + z} \;,
\label{esy}
\end{equation}
\begin{equation}
\eerc = {D^2 \estar \gcr^2 \over 1 + z} \;,
\label{eerc}
\end{equation}
\begin{equation} 
\essc = {D \eb \gcr^4 \over 1 + z} \;,
\label{essc}
\end{equation}
\begin{equation}
\nfnsy = {V_b D^4 \over 4 \pi d_L^2} {4 \over 3} \, c \sigma_T \, u_B
\gcr^2 \, n_e \, \fsp \, \fsy \;,
\label{nfnsy}
\end{equation}
\begin{equation}
\nfnerc = {V_b D^4 \over 4 \pi d_L^2} {4 \over 3} \, c \sigma_T \, \uext
\gcr^2 \, n_e \, \fsp \, \ferc \;, {\rm~and}
\label{nfnerc}
\end{equation}
\begin{equation}
\nfnssc = R_b \, {V_b D^4 \over 4 \pi d_L^2} {4 \over 3} \, c \sigma_T^2
\, u_B \, \gcr^4 \, \left( n_e \fsp \right)^2 \, \fssc \;.
\label{nfnssc}
\end{equation}
The subscripts ``sy," ``SSC," and ``ERC" refer to the synchrotron,
SSC, and external Compton components, respectively, and $\estar$ is the
mean photon energy of the external soft photon field in the stationary
frame of the AGN. The factor $\fsp$ in equations (\ref{nfnsy}) --
(\ref{nfnssc}) is a normalization factor defined through
$\int_1^{\infty} d\gamma \, n_e (\gamma) \, \gamma^2 = \gcr^2 \, n_e
\, \fsp$, and $\fssc$ in equation (\ref{nfnssc}) is a correction
factor between the $\nfn$ peak value and the total energy output in
the SSC component, which differ significantly due to the substantial
spectral broadness of the SSC emission. Typically, $\fssc \sim 0.1$.
For the more strongly peaked synchrotron and ERC components, the
corresponding correction factor is $\fsy \approx \ferc \sim 1$.

Combining the above estimates, and defining $d_L = 10^{28} \, 
d_{28}$~cm and $R_{\rm sc} = 10^{18} \, R_{{\rm sc}, 18}$~cm, 
we find
\begin{equation}
D = 3.2 \, \sqrt{ \eercn \, \esyn \, (1 + z) \over \estarn \, \esscn}\;,
\label{D}
\end{equation}
\begin{equation}
B = 1.4 \, \sqrt{ \esyn^3 \, \estarn \, (1 + z) \over \esscn \, \eercn}\;,
\; {\rm G}
\label{B}
\end{equation}
\begin{equation}
L_D \, \taur  = 3 \times 10^{45} \, 
{\nfnercn \over \nfnsyn} \, {(1 + z) \, R_{{\rm sc}, 18}^2\,\esyn^2 \, \estarn^2 
\over \eercn^2} \; ~{\rm erg~s}^{-1}\;,
\label{LD}
\end{equation}
\begin{equation}
R_{\rm esc} = 1.2 \times 10^{16} \, {\eercn \over (1 + z) \, \estarn} \, 
\sqrt{\esscn \over \esyn^5} \left( 1 + {\nfnercn \over \nfnsyn} \right)^{-1}
\; {\rm cm}\;,
\label{Resc}
\end{equation}
\begin{equation}
R_b = 10^{17} \, {d_{28} \over \sqrt{1 + z}} \, \sqrt{ \nfnsyn^2 \over 
\nfnsscn} \, \sqrt{ \estarn \esscn^3 \over \eercn 
\, \esyn^5} \, \sqrt{(\fssc / 0.1) \over \fsy^2} \; {\rm cm} \; , \;{\rm and}
\label{Rb}
\end{equation}
\begin{equation}
f_j = 4 \, (3 - s) \, {d_L^2 \, \nfnsy \over L_D \, \gamma_1^{s - 2}
\, \fsy} \, \left( {\esy \over \essc} \right)^{-(2 + s)/2} \, \left( {\estar
\over \eerc} \right)^2 \, \left( 1 + {\nfnerc \over \nfnsy} \right).
\label{fj}
\end{equation}
Note that equation (\ref{fj}) requires an independent estimate of
$\gamma_1$.  Assuming that the optical spectrum is produced through
synchrotron emission from strongly cooled electrons, the injection
spectral index $s$ can be estimated through the optical spectral index
$\alpha_{\rm opt}$ from the relation $s = 2 \, \alpha_{\rm opt}$.

The requirement that $\eta = {R_{\rm esc} / R_b} \ge 1$ yields a
constraint on the external soft photon energy $\estar$ from equations
(\ref{Resc}) and (\ref{Rb}), given by
\begin{equation}
\estar \le {2.4 \times 10^{-6} \over \left( \sqrt{1 + z} \, d_{28} 
\right)^{2/3}} \, {\eercn \over \esscn^{2/3}} \, \left( {\sqrt{\nfnsscn} 
\over \nfnsyn + \nfnercn} \right)^{2/3} \left( {\fsy \over 
\sqrt{(fssc/10)}} \right)^{2/3}\;.
\label{estar}
\end{equation}

However, we point out that the observables $\essc$ and $\nfnssc$
are generally uncertain by at least an order of magnitude since the
SSC component does generally not show up as a pronounced bump in the
broadband spectra of FSRQs. Also the ERC peak energy $\eerc$ can rarely 
be determined to an accuracy of better than an order of magnitude. 
Thus, great caution must be used when drawing conclusions from Eq. 
(\ref{estar}).

For the example of 3C~279, one of the best-observed EGRET-detected
FSRQs, we can estimate typical values of the observables as $\esy 
\sim 3 \times 10^{-7}$, $\eerc \sim 300$, $\essc \sim 0.1$, $\nfnsyn
\approx 1$, $\nfnercn \approx 5$, and $\nfnsscn \approx 0.5$
(see, e.g., \cite{hartman01}).  For 3C 279, $z = 0.538$ and
$d_{28} = 0.77$, and we take $\taur = 0.1$, $\gamma_1 = 100$, 
and $s = 2.2$. Equation (\ref{estar}) then implies
$\estar \ls 1.8 \times 10^{-6}$ assuming $\eta = 1$
(corresponding to $E \approx 1$~eV or $\nu \approx 2 \times
10^{14}$~Hz). Possible implications of this soft seed photon 
energy estimate will be discussed in \S \ref{sequence}. Other 
model parameter estimates based on these observables are 
$D = 28$, $B = 2.2$~G, $R_{\rm sc} = 1.2 \times 10^{18} 
\, \sqrt{L_{46} \, (\taur / 0.1)}$~cm, $R_b = 1.7 \times 
10^{15}$~cm, $f_j = 6.4 \times 10^{-4} \, L_{46}^{-1} \, 
(\gamma_1 / 100)^{-0.2}$, implying $L_j \sim 6.4 \times 10^{42}
\, (\gamma_1 / 100)^{-0.2}$~ergs~s$^{-1}$. The high-energy 
cut-off of both the synchrotron and the Compton components 
are poorly known in 3C~279 and other FSRQs, so no reliable
constraint on the maximum electron energy $\gamma_2$ can be 
derived in the FSRQ case.

The parameter estimates for the FSRQ case quoted above are
in good agreement with those found by \cite{ghisellini98} for
FSRQs in general, and through more detailed spectral modeling 
of 3C~279 in particular (e.g., \cite{hartman01}), which indicates 
that our analytic estimates may indeed be useful to assess a 
convenient starting point for more detailed spectral modeling
of FSRQs.

\section{\label{sequence}Evolutionary Sequence from FSRQs to HBLs}

In the previous section, we obtained typical parameter values for HBLs
and FSRQs. These estimates indicate that the most significant
differences between model parameters of FSRQs and HBLs are found in
the values of the jet power $L_j$ and the magnetic field $B$,
which are both found to decrease by more than an order of magnitude
from FSRQs to HBLs. The lack of evidence for a significant
contribution of an external Compton component to the $\gamma$-ray
spectra of HBLs indicates that the energy density of the external soft
photon field --- parameterized through $L_D$, $\taur$, and $R_{\rm
sc}$ --- decreases by $\gs 3$ orders of magnitude along the
sequence of blazar types. This is in agreement with the results
of \cite{ghisellini98}.

The decreasing energy density of the external soft photon field
along the sequence FSRQ $\to$ LBL $\to$ HBL, in tandem with the
decreasing jet power suggests that this is a consequence of a
decreasing accretion rate along this sequence. This, in turn, is
in very good agreement with the suggestion of \cite{delia00} and
\cite{cavaliere01} of an evolutionary sequence from FSRQs to BL Lacs 
characterized by such a decline in the accretion rate. We are thus 
motivated to try to reconstruct the evolutionary sequence FSRQ $\to$ 
LBL $\to$ HBL by varying the accretion rate (and other model parameters 
dependent on the accretion rate) in our specific spectral blazar model 
described in the previous section. In order to do so, we are starting 
out with parameters appropriate to reproduce the broadband spectrum of 
a typical FSRQ. The FSRQ phase constitutes an early phase of
blazar evolution in which the central regions of the galaxy are rich
in gas and dust, leading to a high accretion rate onto the central,
supermassive black hole. At the same time, the circumnuclear material
efficiently reprocesses and scatters the accretion-disk radiation,
leading to the observed strong optical emission lines in the broad
line region and to a high energy density of the external soft photon
field in the jet. 

Due to the limited supply of matter in the nuclear region, the 
average density of the circumnuclear material will gradually 
decrease, leading to a decreasing accretion rate and a decreasing 
reprocessing efficiency. We parametrize this evolutionary transition
by a gradual decline of the accretion rate and assume that 
the reprocessing efficiency $\taur$ is proportional to the
accretion rate. This should be a reasonable approximation for the
long-term evolution on cosmological time scales, although the
actual BLR density and the instantaneous accretion rate may be
less strictly correlated on shorter time scales. Furthermore, we
assume that the fraction of accreted energy channeled into the 
relativistic outflow, $f_j$, remains constant, as suggested by
the observed $L_j$ vs. $L_{\rm BLR}$ and $L_j$ vs. $L_{\rm NLR}$
correlations \citep{rs91,celotti97,xu99}, and that the transverse
extent of the jet, $R_b$, does not change dramatically during this
evolution. The magnetic field is chosen to be a constant fraction 
of the equipartition magnetic field so that $u_B = 0.1 \, u_e$,
which automatically leads to the required reduction of the magnetic 
field along the evolutionary sequence.

\begin{figure}
\plotone{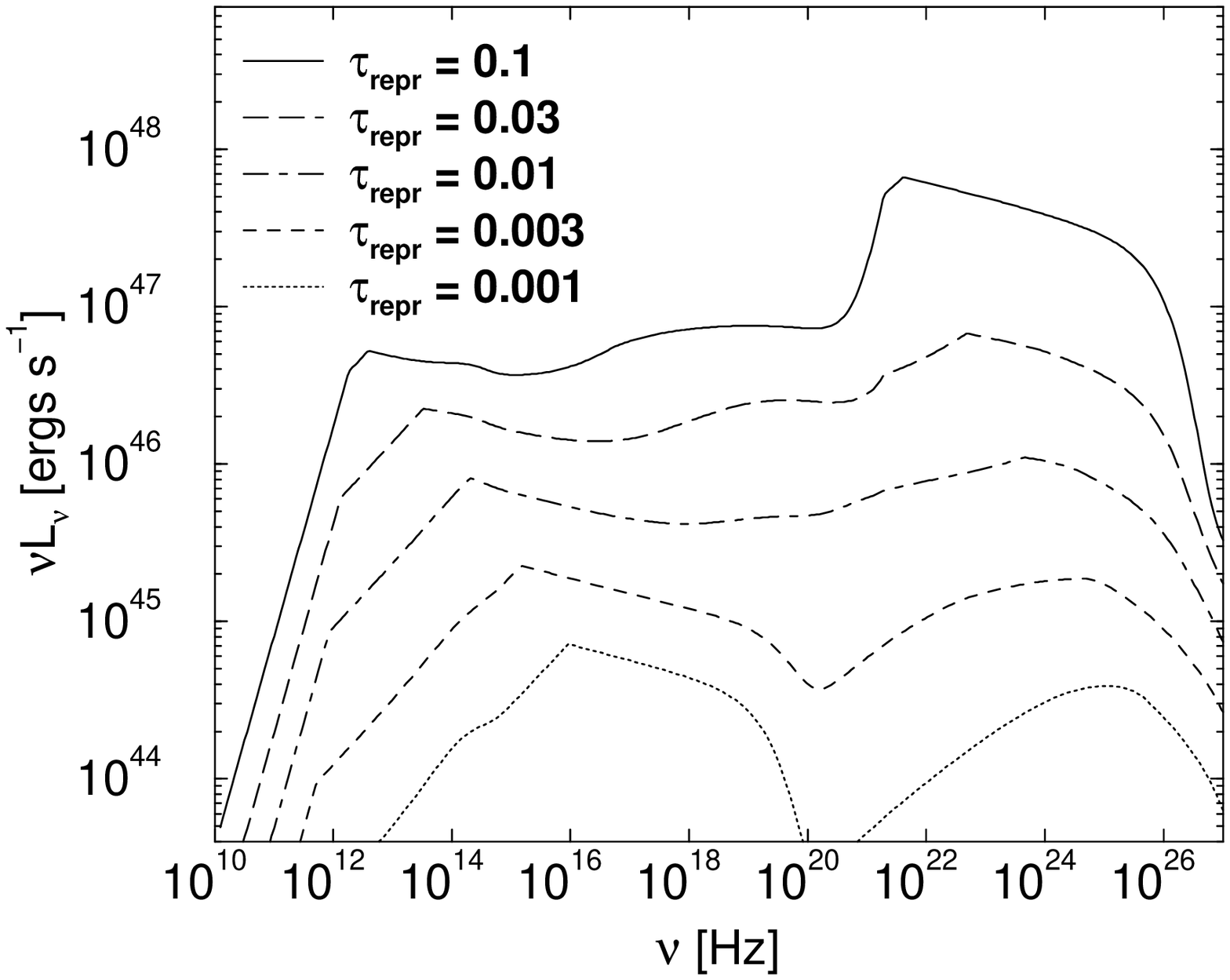}
\caption{One-parameter model sequence of broadband blazar spectra that 
reproduces the trend in the spectral energy distributions of FSRQs,
LBLs, and HBLs. Starting with typical FSRQ parameters, $\gamma_1 =
100$, $\gamma_2 = 10^6$, $s = 2.2$, $D = 28$, $R_b = 3 \times 10^{15}$~cm,
$B = 2.4$~G, $L_D = 10^{46}$~ergs~s$^{-1}$, $\taur = 0.1$, $\estar = 
10^{-6}$, $f_{jet} = 10^{-3}$, $R_{\rm sc} = 1.2 \times 10^{18}$~cm, 
the sequence is generated by reducing $\taur$ and assuming that the 
accretion disk luminosity is proportional to $\taur$. The comoving 
magnetic field $B = 0.3 \, B_{\rm ep}$, where $B_{\rm ep}$ is the 
equipartition magnetic field with electrons.}
\label{fig_sequence}
\end{figure}

The resulting sequence of broadband spectra, shown in Fig.\ 
\ref{fig_sequence}, provides a quantitative explanation for 
the observed trend of luminosities and peak photon energies 
in the FSRQ $\to$ LBL $\to$ HBL sequence \citep{fossati98}.  
It does not account for the spectral properties of specific
blazars, which depend on variations of parameter values in
different sources and, importantly, the angle between the direction of
the jet axis and the direction to the observer. We note that the
apparent minimum at $\nu \sim 10^{20}$~Hz in the FSRQ spectrum
(corresponding to the $\taur = 0.1$ curve in Fig.\ \ref{fig_sequence}) 
might be filled in by an additional contribution from 
Compton-scattered accretion disk radiation that enters 
the jet directly, which has been neglected in our 
semi-analytic model.

Our estimate (Eq.\ [\ref{estar}]) for the average energy 
$\estar \approx 2 \times 10^{-6}$ of the dominant external soft 
photon field has interesting implications. Although one 
has to keep in mind that the analytical estimate for $\estar$
may be rather uncertain due to the uncertainty in the observables
pertaining to the SSC component, we found that significantly higher
values of $\estar$ did not allow an acceptable representation of
a typical FSRQ spectrum with our model. This probably indicates 
that the external soft photon field is dominated by optical line 
emission from the partially ionized broad-line regions. The external
photon field peaking at $\estar$ is unlikely to be due to Thomson 
scattering off free electrons in a highly ionized circumnuclear 
environment, since in that case $\estar$ should reflect the peak 
energy of the optically thick accretion disk, which would be 
inconsistent with the evidence for thermal excess emission, 
peaking at $\sim 2 \times 10^{15}$~Hz, observed in the low 
$\gamma$-ray state of 3C~279 by \cite{pian99} and with the
peak frequency of $\sim 3 \times 10^{15}$~Hz of the big blue 
bump seen in 3C~273 \citep{lichti95} which is generally 
attributed to the emission from the optically thick accretion 
disk. 

\section{\label{discussion}Discussion and Summary}

The proposed evolutionary scenario provides a simple physical
connection between different blazar subclasses in terms of the
depletion of the accretion flow onto supermassive black holes.  After
an early phase of rapid black-hole growth, the time scale for this
depletion is given by the matter free-fall time $t_{\rm ff} \sim 7
\times 10^8 \, m_8 \, T_4^{-3/2} \, {\rm yr}$ at the Bondi-Hoyle
accretion radius, where $m_8$ is the mass of the central black hole
in units of $10^8\, M_{\odot}$ and $T = 10^4 \, T_4$~K is the average 
temperature of the circumnuclear material. This is generally shorter 
than the Hubble time, and gives a plausible time scale to explain a 
finite duration of blazar-like activity in the nuclei of elliptical
galaxies. The scenario proposed here is also in accord with the X-ray
faintness of giant elliptical galaxies known to contain supermassive
black holes, which may be remnants of past blazar activity. The
central sources in these galaxies have been successfully modeled
with advection-dominated accretion flow solutions at very low accretion 
rates, $l\ll 1$ \citep{fr95,dmf97,dm00}, suggesting a central region 
depleted after an earlier period of more efficient accretion.

If this picture is correct, then there should also be objects in an
earlier phase of rapid black-hole growth that are accreting at rates
$l \sim 1$. Following the scenario outlined by \cite{san88}, this
earlier stage of blazar evolution would comprise merging galaxies,
infrared luminous galaxies, and radio-quiet QSOs. Accretion flows near
the Eddington limit might produce optically-thick accretion disks
extending all the way to the innermost stable orbit. If the
phenomenology of Galactic black-hole candidates can be scaled to
supermassive black holes accreting at comparable values of $l$, then
radio jet formation should be quenched in such a mode of accretion as
is observed, e.g., in GRS~1915+105
\citep{mirabel98,feroci99,belloni00}, GX~339-4
\citep{fender99,corbel00}, and XTE J1550-564 \citep{corbel01}. 
This idea is supported by recent theoretical work
that links the existence and energetics of a central
advection-dominated inflow-outflow system to the formation of
relativistic jets \citep{bb99,becker01}.  The ASCA detections of
relativistically broadened Fe K$\alpha$ emission in MCG-6-30-15
\citep{tanaka95} and IRAS~18325-5926
\citep{iwas96} and the rapid X-ray variability seen 
in IRAS~13224-3809 and PHL~1092 \citep{boller00}, suggest that
at least in a number of cases an optically thick accretion 
disk does extend towards the innermost stable orbit. In most 
cases the evidence is yet inconclusive, but a similar mode of 
accretion can generally not be ruled out.

The formation of radio jets in blazars and radio galaxies, if related
to the formation of an advection-dominated accretion mode in the inner
portions of the accretion flow, is triggered by a decreasing Eddington
ratio. The decline of $l$ might be due to a combination of a
decreasing accretion rate and an increasing black-hole mass. An
implication of this scenario is that the masses of the central black
holes in galaxies which host BL Lacs should, on average, be greater
than the masses of black holes in galaxies which host FSRQs or optical
QSOs. Moreover, subclasses at earlier stages in the blazar sequence
should exhibit increasingly stronger cosmological
evolution. \cite{bad98} find evidence for negative cosmological
evolution of X-ray selected BL Lac objects, and \cite{sti91} find
evidence for positive cosmological evolution in a radio-selected BL
Lac sample, in accord with this picture.  Our work provides strong
support for the suggestion by \cite{delia00} and \cite{cavaliere01} along 
parallel lines, but on the basis of general energy considerations and a 
generic cosmological evolution scenario for quasars evolving into BL Lac 
objects. We have combined the basic idea of such a cosmological evolution 
with a self-consistent spectral blazar model and demonstrated that the 
spectral properties of the different blazar subclasses can be consistently 
explained by such an evolutionary sequence. The blazar model adopted here
is similar to the one previously used by \cite{ghisellini98} --- but with 
the important addition of particle escape and adiabatic losses --- who 
had argued that the different blazar subclasses could be unified by an
increasing level of external radiation to the soft radiation field in
the emitting region.

To summarize, we have proposed an evolutionary scenario linking FSRQs,
LBLs, and HBLs through gradual depletion of the circumnuclear
environment of a supermassive black hole.  As the accretion power
declines with time, less gas and dust is available to reprocess
accretion-disk radiation and produce an external Compton-scattered
component in the blazar jet. Analytic formulae are given to estimate
the relevant model parameters for FSRQs. We have calculated a spectral
sequence for different blazar subclasses that are related through a
one-parameter model defined by the optical depth of the circumnuclear
matter.  This scenario links radio-loud AGNs with progenitor merger
galaxies and QSOs that constitute the high Eddington-ratio limit of
the evolutionary sequence. In analogy with the observed
anti-correlation between radio and soft X-ray activity in some
Galactic black-hole candidates, these progenitor sources would accrete
with large Eddington ratios, resulting in an optically-thick accretion
disk that quenches relativistic jet production.

\acknowledgments
We thank the anonymous referee for a very careful review and helpful
and constructive comments. The work of MB is supported by NASA through 
Chandra Postdoctoral Fellowship grant PF~9-10007 awarded by the Chandra 
X-ray Center, which is operated by the Smithsonian Astrophysical 
Observatory for NASA under contract NAS~8-39073. The work of CD is 
supported by the Office of Naval Research.

\newpage

\end{document}